\documentclass{elsart}

\usepackage{amssymb}

\begin{document}

\begin{frontmatter}

\title{Superconducting fluctuations in small grains - the Universal Hamiltonian and the reduced BCS
model}

\author{Moshe Schechter}

\address{Department of Physics \& Astronomy, University of British Columbia,
Vancouver, B.C., Canada, V6T 1Z1}

\begin{abstract}
Small superconducting grains are discussed in the frameworks of both
the reduced BCS Hamiltonian and the Universal Hamiltonian. It is
shown that fluctuations of electrons in levels far from the Fermi
energy dominate superconducting properties in small and ultrasmall
grains. Experimental consequences related to the spin susceptibility
and persistent currents of grains and rings with weak
electron-electron interactions are discussed.
\end{abstract}

\end{frontmatter}

\section{Introduction}
\label{introduction}

The paradigmatic framework in the study of superconducting
fluctuations in small grains\cite{DR01} is the reduced BCS
Hamiltonian

\begin{equation}
\hat{H} = \sum_{j,\sigma=\pm} \epsilon_j c^\dagger_{j\sigma}
c_{j\sigma} - \lambda d \sum_{i,j}  c^\dagger_{i+} c^\dagger_{i-}
c_{j-} c_{j+} . \label{BCS}
\end{equation}
Here $\lambda$ is the dimensionless interaction constant, $d$ is the
mean level spacing, and the indices $i,j$ correspond to doubly
degenerate time reversed states of energy $E_{\rm F} - \omega_{\rm
D} < \epsilon < E_{\rm F} + \omega_{\rm D}$.

While the mean field (BCS)\cite{BCS57} solution of this Hamiltonian
was extremely powerful in explaining the properties of bulk
superconductors, grains of finite size pose additional challenges.
In such grains the number of electrons is fixed, their size may be
smaller than the coherence length $\xi$ and the magnetic penetration
depth, and zero resistance is not achievable. Still, already in
$1959$ Anderson postulated that superconducting grains maintain
superconducting properties for sizes much smaller than $\xi$, down
to the size in which the single electron level spacing $d$ equals
the bulk energy gap $\Delta$ (the "Anderson size"\cite{And59}). For
clean grains this size is of order $\xi^{1/3} \lambda_{\rm F}^{2/3}
\ll \xi$.

Much of the recent theoretical interest in the study of small
superconducting grains (see Ref.\cite{DR01} and references therein)
was initiated by the experimental work of Ralph Black and
Tinkham\cite{BRT96}, who found a parity dependent gap in the
excitation spectrum of single "small" Al grains (with $d<\Delta$),
but not for "ultrasmall" grains having $d>\Delta$. More recently it
was also shown that the Meissner effect sharply disappears at grain
sizes consistent with the Anderson size\cite{RLPS03,LYTL03,BRB+05}.

While superconducting characteristics reminiscent of bulk properties
indeed vanish at the Anderson size, signatures of pairing
correlations persist to smaller sizes. These signatures are a result
of superconducting fluctuations of electrons at energies larger than
the gap energy, up to the cutoff energy given by $\omega_{\rm D}$.
For example, the spin susceptibility of ultrasmall superconducting
grains is predicted to exhibit a re-entrant behavior both as
function of temperature\cite{LFH+00} and as function of magnetic
field\cite{SILD01,Sch04}, with a long tail, persisting up to the
temperature/magnetic field equivalent of $\omega_{\rm D}$. The
superconducting fluctuations of electrons further than $\Delta$ from
the Fermi energy $E_{\rm F}$ affect significantly also grains in an
intermediate regime, where $\Delta^2/\omega_{\rm D} < d < \Delta$.
In particular, in this regime the energy gain of the system by the
attractive interaction is much larger than that given by the mean
field treatment of BCS\cite{SILD01}, and the same is true for the
pairing parameter\cite{SDIL03}. Moreover, it was shown that these
far level fluctuations affect differently single particle and
collective properties, and therefore one has to define
correspondingly two different pairing parameters\cite{SDIL03} when
discussing small superconducting grains.

While the above properties, and specifically their dependence on the
high energy cutoff, stem from the exact solution of the reduced BCS
hamiltonian\cite{Ric63,RS64}, the question to their relevance to
real superconducting grains may arise\cite{YBA05}. This question is
of particular relevance since specific predictions for the
experimental consequences of the fluctuations of the far levels
(e.g. the long tail of the spin susceptibility discussed above) are
made. Here we discuss the significance of the superconducting
fluctuations of the far levels by considering both the reduced BCS
model and the model of the Universal hamiltonian\cite{KAA00}. While
the reduced BCS Hamiltonian is an effective Hamiltonian whose
validity in describing high energy properties may be questioned, the
Universal Hamiltonian was shown, using renormalization group
approach, to control the low energy physics of metallic grains with
weak interactions and large dimensionless
conductance\cite{MM02,MS03}. We thus establish the significance of
the superconducting fluctuations of the far levels up to the high
energy cutoff of the Universal Hamiltonian given by the Thouless
energy $E_{\rm Th}$. We then discuss the significance of the Debye
energy in general, and for the problem of persistent currents in
particular.

\section{Reduced BCS Hamiltonian and Universal Hamiltonian - exact
solution} \label{sec2}

The reduced BCS Hamiltonian (\ref{BCS}), with finite number of
electrons, was solved exactly by Richardson\cite{Ric63,RS64}. The
structure of the solution takes advantage of the fact that the
interaction scatters only pairs and not single electrons. Thus, the
solution is obtained separately for subspaces of the Hilbert space
defined by the identity of the singly occupied levels. The singly
occupied levels are neglected at first, the problem of $N$ pairs in
$M$ states is solved, and the singly occupied levels are then
trivially added to the solution (see Refs.\cite{Ric63,RS64} as well
as Ref.\cite{DB00} for details).

While the reduced BCS Hamiltonian is motivated by the specific
phonon-mediated attractive electron-electron interaction, it was
recently shown that any metallic grains with large dimensionless
conductance $g \equiv E_{\rm Th}/d$, weak interactions, and
negligible spin-orbit interaction, can be described by the Universal
Hamiltonian\cite{KAA00}, which includes only three interaction
parameters

\begin{equation}
H = \sum_{n,\sigma} \epsilon_n c^{\dagger}_{n,\sigma} c_{n,\sigma} +
E_c \hat{N}^2 + J_c \hat{T}^{\dagger} \hat{T} + J_s \hat{S}^2.
\label{universal}
\end{equation}
Here $\hat{N} = \sum_{n,\sigma} c^{\dagger}_{n,\sigma} c_{n,\sigma}$
is the number operator, $\hat{\vec{S}} =\frac{1}{2} \sum_{n,\sigma,
\sigma '} c^{\dagger}_{n,\sigma} \vec{\sigma}_{\sigma,\sigma '}
c_{n,\sigma '}$ is the total spin operator, and $\hat{T} = \sum_{n}
c_{n,-} c_{n,+}$ is the pair annihilation operator. The index $n$
spans a shell of doubly degenerate time reversed states of energy
$E_{\rm F} - E_{\rm Th} < \epsilon_n < E_{\rm F} + E_{\rm Th}$.
$E_c$ is the charging energy and $J_{c(s)} = \lambda_{c(s)} d$,
where $\lambda_c$ and $\lambda_s$ are the dimensionless interaction
parameters in the Cooper channel and in the spin channel
respectively.

Interestingly, for isolated grains the solution of the Universal
Hamiltonian is given by Richardson's solution for the reduced BCS
Hamiltonian. First, the Coulomb term can be neglected, since the
number of electrons in the isolated grain remains constant. Then,
other than a different cutoff energy ($E_{\rm Th}$ compared to
$\omega_{\rm D}$), one remains with the reduced BCS Hamiltonian with
the additional exchange term. The latter commutes with the rest of
the Hamiltonian. Moreover, the pairing interaction involves solely
the paired levels, while the spin interaction involves solely the
singly occupied levels. One can then obtain the solution for the
Universal Hamiltonian by following the steps in Richardson's
solution of the reduced BCS Hamiltonian, with the additional
consideration of the spin term for the singly occupied levels.

\section{Large contribution of the far levels}

Considering the reduced BCS Hamiltonian (\ref{BCS}), and using
Richardson's exact solution, the condensation energy of small
metallic grains was calculated as function of the coupling constant
$\lambda$\cite{SILD01}. The condensation energy was defined as the
difference between the energy of the Fermi state and the real ground
state of the system, i.e.

\begin{equation}
E_{\rm cond}(\lambda) \equiv E_{\rm F.g.s}(\lambda) - E_{\rm
g.s.}(\lambda).
\end{equation}
For bulk superconductors this energy is given by $\Delta^2/(2 d)$,
is extensive, and is a non-analytic function of $\lambda$. For
finite size grains it was found that the condensation energy is
analytic at $\lambda=0$, and is very well estimated by $E_{\rm cond}
\simeq \Delta^2/(2 d) + \ln{2} \lambda^2 \omega_{\rm D}$. The first
term gives the contribution of the levels within $\Delta$ of $E_{\rm
F}$, and the second, perturbative term, is the contribution of the
far levels. Interestingly, due to unique analytical properties of
the condensation energy, the perturbative term expresses the
contribution of the far levels not only within the regime of
validity of perturbation theory ($\lambda < 1/\ln{N}$, or
$d>\Delta$), but also in the regime where $d<\Delta$\cite{SILD01}.
An immediate outcome of this result is that for a large intermediate
regime, $\Delta^2/\omega_{\rm D} < d < \Delta$, the condensation
energy is much larger than that given by the BCS term. Note that the
perturbative term is intensive, and therefore negligible as the size
of the grain becomes large.

Similarly, the contribution of the far levels is found to be
significant when considering generalizations of the bulk order
parameter that are suitable for finite size grains\cite{SDIL03} (see
also Ref.\cite{DR01} and references therein). For all standard
definitions of the order parameter the far level contribution
results in a term linear in $\omega_{\rm D}$, and the order
parameter being much larger than its mean field value in the
intermediate regime defined above (see Ref.\cite{SDIL03} for
details). Thus, in contrast to the common belief that the order
parameter turns from being extensive to being intensive at the
Anderson size ($d \approx \Delta$), it actually becomes intensive
already at a much larger size, i.e. when $d \approx
\Delta^2/\omega_{\rm D}$. Furthermore, the significant contribution
of the far levels separates collective properties of the
superconductor such as the condensation energy, from single particle
properties such as the energy gap for single particle excitations.
While in bulk superconductors both energies are related to the bulk
order parameter $\Delta$, for small grains one has to define two
different order parameters to describe collective and single
particle properties\cite{SDIL03}. While the cutoff energy affects
the former linearly, it affects single particle properties only
logarithmically.

Interestingly, using the similarity of the reduced BCS and the
Universal Hamiltonians, and their exact solution, the results
discussed above for the reduced BCS Hamiltonain are immediately
applicable to the Universal Hamiltonian, with the only change of the
cutoff from $\omega_{\rm D}$ to $E_{\rm Th}$. Thus, as long as the
grains obey the conditions of applicability of the Universal
Hamiltonian described above, one obtains the intermediate regime,
defined now by $\Delta^2/E_{\rm Th} < d < \Delta$, in which e.g. the
condensation energy is much larger than the mean field BCS value,
and therefore intrinsic. The predictions of re-entrant spin
susceptibility as function of temperature\cite{LFH+00} and magnetic
field\cite{SILD01,Sch04}, which result from the pairing correlations
of the far level, can also be made on the basis of the validity of
the Universal Hamiltonian. Importantly, all the above
characteristics do not depend on the interaction being constant up
to the upper cutoff, nor do they depend on the existence of a sharp
cutoff. It is sufficient that one can bound the interaction from
below by $c_1 \lambda_c$ within a window of $c_2 E_{\rm Th}$ from
$E_{\rm F}$, where $c_1,c_2$ are constants of order unity. Thus,
based on the validity of the Universal Hamiltonian one can argue for
the applicability of the above results for the condensation energy
and spin susceptibility for real physical systems. Note, that within
the regime of applicability of the Universal Hamiltonian one can
uniquely relate the excess spin susceptibility as function of the
magnetic field to the existence of pairing correlations, as the
exchange term can not account for such behavior. Furthermore, by
examining the behavior of the spin susceptibility of an ensemble of
small metallic grains the interaction parameters of the Universal
Hamiltonian can be determined\cite{Sch04}. Another consequence of
the above analysis is that the interesting question of whether the
noble metals have a weak attractive interaction and therefore are
superconductors albeit with a very low $T_c$ can be addressed as
well, in small grains and by extrapolation in bulk. This is since
unlike $T_c$ which is exponentially small in $\lambda_c$, the excess
spin susceptibility as a function of the magnetic field is quadratic
in $\lambda_c$ and therefore detectable for small $\lambda_c$.

\section{Persistent currents and the significance of $\omega_{\rm
D}$}

In the discussion above it is argued that the interesting physics
related to the pairing correlations of the far levels is independent
from the validity of the reduced BCS model in describing the
properties of small grains at energies of the order of $\omega_{\rm
D}$. Still, the energy scale of $\omega_{\rm D}$ is a physical
energy scale, related to the retardation of the phonon mediated
interaction. Especially since, as was mentioned above for the
Universal Hamiltonian, a constant interaction and a sharp cutoff at
$\omega_{\rm D}$ are not necessary for the applicability of the
above results, it is plausible that indeed the energy scale of
$\omega_{\rm D}$ dictates the magnitude of the condensation energy
of small grains, the boundary of the intermediate regime, and the
extent of the tail in the excess spin susceptibility.

However, in relation to persistent currents, a rigorous
understanding of the interaction is even more crucial. Since the
persistent current in a normal metal ring is given by the derivative
of the energy with respect to the flux, it is plausible that for
metals with weak attractive interaction, the large contribution of
the far levels to the condensation energy will affect the persistent
current, and will result, within the reduced BCS model, in a large
persistent current related to the energy scale of $\omega_{\rm D}$.
Indeed, in Ref. \cite{SOIL03} it was shown that the value of the
derivative of the persistent current at zero flux is much larger
within the reduced BCS model in comparison to the value obtained
within the standard theory of momentum independent
interaction\cite{AE90b}. Here, however, the difference in magnitude
of the persistent current is related to the high energy cutoff being
$\omega_{\rm D}$ rather than $E_{\rm Th}$, and thus is crucially
related to the specifics of the model considered, and to the exact
form of the interaction. In particular, the BCS interaction assumes
that only time reversed states interact, i.e. that the total
incoming momentum of the scattered electrons $q$ must be zero. While
this is a simplified form of the momentum dependence of the
interaction, the subsequent result of the much larger magnetic
reponse\cite{SOIL03} in comparison to the value obtained within the
momentum independent picture points to the importance of taking
correctly the dependence of the interaction on $q$, especially since
a significant $q$ dependence of the attractive interaction is
motivated by the retardation of the phonon mediated
interaction\cite{SOIL03,SOIL04}. We believe that a rigorous
understanding of the form of the interaction, and in particular its
$q$ dependence, could lead to a better understanding of the long
standing question regarding the value of the ensemble averaged
persistent current in small metallic rings.

This work was supported by NSERC of Canada and PITP.

\end{document}